\documentclass[aps,prl,twocolumn,superscriptaddress,10pt]{revtex4-2}
\usepackage{amssymb,amsthm,amsmath,amsfonts}
\usepackage{graphicx,ulem,enumerate,mathptmx}
\usepackage{soul}
\usepackage[pdftex,dvipsnames,usenames]{xcolor}
\usepackage[colorlinks=true,urlcolor=blue,citecolor=blue,linkcolor=blue]{hyperref}
\usepackage{tcolorbox}

\newcommand{\NP}[1]{\textcolor{black}{#1}}

\begin{document}

\title{Photonic Entanglement and Polarization Nonclassicality: Two Manifestations, One Nature}

\author{Laura Ares}
\email{laurares@mail.uni-paderborn.de}
\affiliation{Theoretical Quantum Science, Institute for Photonic Quantum Systems (PhoQS), Paderborn University, Warburger Stra\ss{}e 100, 33098 Paderborn, Germany}

\author{Nidhin Prasannan}
\affiliation{Integrated Quantum Optics Group, Institute for Photonic Quantum Systems (PhoQS), Paderborn University, Warburger Stra\ss{}e 100, 33098 Paderborn, Germany}

\author{Elizabeth Agudelo}
\affiliation{Atominstitut, Technische Universit\"at Wien, Stadionallee 2, 1020 Vienna, Austria}

\author{Alfredo Luis}
\affiliation{Departamento de  \'Optica, Facultad de Ciencias F\'{\i}sicas, Universidad Complutense, 28040 Madrid, Spain}

\author{Benjamin Brecht}
\affiliation{Integrated Quantum Optics Group, Institute for Photonic Quantum Systems (PhoQS), Paderborn University, Warburger Stra\ss{}e 100, 33098 Paderborn, Germany}

\author{Christine Silberhorn}
\affiliation{Integrated Quantum Optics Group, Institute for Photonic Quantum Systems (PhoQS), Paderborn University, Warburger Stra\ss{}e 100, 33098 Paderborn, Germany}

\author{Jan Sperling}
\affiliation{Theoretical Quantum Science, Institute for Photonic Quantum Systems (PhoQS), Paderborn University, Warburger Stra\ss{}e 100, 33098 Paderborn, Germany}
       
\date{\today}

\begin{abstract}
    We demonstrate in theory and experiment the strict equivalence between nonclassical polarization and the entanglement of indistinguishable photons, thereby unifying these two phenomena that appear dissimilar at first sight.
    This allows us to analyze nonclassicality and multi-photon entanglement within the same framework.
    We experimentally verify this double-sided form of quantumness and its independence from the polarization basis, contrasting other notions of coherence that are highly basis-dependent.
    Our findings show how nonclassical polarization turns out to be equally resourceful for quantum protocols as entanglement, emphasizing its importance in practical applications.
\end{abstract}
\maketitle

\paragraph{Introduction.}

Today, we recognize a plethora of behaviors that cannot be understood via classical theories, having a place entirely in the quantum framework \cite{HHHH2009, AGML2016, FVMH2018, SV2020}. 
Grasping these effects not only deepens our knowledge but is vital for developing technologies. 
Consequently, it remains crucial to examine the boundary between the quantum and classical domains \cite{S2016, KL2019, MKKW2019, K2023, BF2023}, as well as between the different forms of quantum correlations that exist in nature \cite{FP2012,ASV2013,KASFSSVH2021}. 
Nowadays, we take advantage of quantum effects to considerably improve the performance of tasks, such as communication, simulation, computation, and sensing \cite{LKNBLGF2018, GHKPU2023, PAALBOCETLOPUVWVW2020, R2020, DBKFPTZ2022, BDGJGN2022}.
Such nonclassical traits are often investigated as resources that must be optimized, looking for the most effective performance possible \cite{CG2019}. 

Most quantum features are direct consequences of the superposition principle and, therefore, depend on the chosen representation \cite{FHBWLSG2021}, implying that certain correlations can be transformed into different ones by means of classical operations \cite{AL2021, VS2014}. 
This ambiguity fuels the investigation of the differences and the shared essence of nonclassical features \cite{KASFSSVH2021, GGLSS2022, AL2022}. 
Still, it remains unclear whether any of these concepts may assume the role of properly unifying certain notions of quantumness. 
Here, we focus on the common understanding of two outstanding quantum phenomena: entanglement and nonclassical polarization, both being widely investigated due to their usefulness in photonic quantum technologies. 

Without a doubt, entanglement stands out as a quintessential and universally acknowledged quantum phenomenon \cite{HHHH2009}. 
It serves as a prime illustrator for how composite quantum systems and their interference challenge the classical understanding of nature \cite{EPR1935, S1935, S1936}.
It is often considered the key resource for quantum communication protocols \cite{PPHMBT2022} and paramount for advantageous quantum metrology and computing applications \cite{OL2020, A_etal2019}. 
While bipartite qubit systems are undoubtedly the most extensively studied to date, multi-level systems present an even more abundant diversity of high-dimensional entanglement (cf. e.g. \cite{EKZ2020, CVSGND2023, BHKLENBKHBUP2023}), boasting broader encoding alphabets and becoming more resilient to noise.
In addition to the dimensions of each subsystem, increasing the number of parties further amplifies the distinct kinds of quantum correlations.
However, as the complexity grows, the available strategies for certification become increasingly limited.
For these reasons, there are several approaches to detect and quantify the presence of entanglement, among which, the witnessing approaches are particularly effective \cite{VPRK1997, HMGH2010, GT2009, SV2013, CS2014, FVMH2018, MHT2023}.

The second quantum property addressed in this work is nonclassical polarization \cite{L2016,KLLTS2002}. 
As a degree of freedom, polarization is one of the most commonly used attribute for encoding quantum information. 
It is not the only true discrete degree of freedom when quantized fields are under consideration, but the strategies to handle and measure it are readily available \cite{M2012, SBKSL2012, SAJBLB2013}. 
The spectral discreteness of Stokes operators and the unavoidable quantum fluctuations give rise to several nonclassical properties, such as squeezing and hidden polarization \cite{PSBS2022,GHGBKGLS2021}. 
These properties are valued resources for quantum metrology, where polarization squeezing is harnessed to surpass the standard quantum limit \cite{GSYL1987, FAKC2021}.
How to detect and quantify polarization nonclassicality is also a recurrent question that still does not have a definitive answer as it can be addressed via higher-order correlations \cite{USTB2001, GKDLAS2022}, squeezing criteria \cite{KU1993, KIAKL2012}, negative quasiprobability distributions \cite{SCK2017}, etc. 
Strong relationships between entanglement and nonclassical polarization have been found since the very definition of spin-squeezing \cite{KU1993}. 
The squeezing parameter has been directly related to particle entanglement \cite{SDCZ2001, FG2020}, as well as to mode entanglement \cite{KLLRS2002, LK2006, VS2014, SA2023}.

In this letter, we present the theoretical framework and the experimental certification of a unified description for multiphoton entanglement and nonclassical polarization.
We go beyond the mere relation and bring them together via a joint characterization. 
The equivalence relies on the separability of angular-momentum coherent states, and these states turn out to be suitable classical references for both features.
We report on the reconstruction of the quasiprobability distribution of an entangled pair of photons produced by parametric down-conversion.
We obtain significant negativities accounting for the highly nonclassical character of the polarization state and its entanglement.

\paragraph{Angular-momentum coherent states separability.}

In order to simultaneously describe entanglement and nonclassical polarization, we begin with establishing the classical references for each quantum feature separately.
Then we proceed to demonstrate how the description of angular-momentum coherent states in two-mode Fock basis is equivalent to the expansion in polarization basis for $N$ identical particles.
Finally, we show that the quantum effects that share the same set of classical states are effectively going to be characterized by the same quasiprobabilities.

Regarding entanglement, we consider the set of separable states for a $N$-particle system $\{|\psi_1\rangle \otimes \dots \otimes |\psi_N\rangle\}$ as the classical reference \cite{SV2009}. 
Since we are interested in indistinguishable bosons, where $|\psi_1\rangle = \dots = |\psi_N\rangle$ \cite{BFFM2020}, such separable states can be written as $|\psi_1\rangle \otimes \dots \otimes |\psi_N\rangle = |\psi\rangle^{\otimes N}$. 

For polarization, the classical reference is the set of angular-momentum coherent states \cite{ACGT1972}, being regarded as the most classical states in finite-dimensional systems \cite{GBB2008, GBB2010}. 
This label is ascribed because of their minimum uncertainty in the phase space, their invariance under $\mathrm{SU}(2)$ transformations, and their straightforward relation to Glauber coherent states \cite{L2016, GHGBKGLS2021, AD1971}. 
Angular-momentum coherent states, $|\vec{s}\rangle$, can be expanded as the tensor product of indistinguishable photons $|q\rangle$, 
\begin{equation}\label{eq:sq}
|\vec{s}\rangle \leftrightarrow |q\rangle^{\otimes N}.
\end{equation}
To see this, we start from their general expansion in the two-mode Fock basis $\{|n_H , n_V \rangle\}$, 
\begin{equation}
    \label{eq:vecs2}
    |\vec{s}\rangle = \sum_{\substack{n_H,n_V=0 \\ n_H+n_V=N}}^N \binom{N}{n_H}^{1/2}\chi^{n_H}\xi^{n_V}|n_H,n_V\rangle,
\end{equation}
with  $|n_H , n_V \rangle = |n_H \rangle \otimes | n_V \rangle$.
Furthermore, the polarization state of a single photon can be expressed in the basis formed by the horizontal and vertical polarization states, $\{|H\rangle, |V\rangle\}$.
Therefore, any pure state (or qubit) becomes $|q\rangle = \alpha_H|H\rangle + \alpha_V|V\rangle$, with $|\alpha_H|^2 + |\alpha_V|^2 = 1$. 
Then, the tensor product of $N$ indistinguishable photons reads as follows:
\begin{align}
    \label{eq:qn}
    |q\rangle^{\otimes N}=\displaystyle\sum_{m=0}\alpha_H^m \alpha_V^{N-m} (|H\rangle^{\otimes m} |V\rangle^{\otimes N-m}+\mathrm{permutations}).
\end{align}

States in Eq. \eqref{eq:qn} belong to a $2N$-dimensional Hilbert space, $\mathcal{H}^{\otimes N}$, whose basis can be defined by the tensor product of the bases of $N$ two-dimensional Hilbert-spaces $\{|k_1\rangle \otimes \dots \otimes |k_N\rangle\}$ for $k_1,...,k_n\in\{H, V\}$.
The key point of this derivation is the connection that can now be made between the elements of these bases.
Specifically, the states composing the two-mode Fock basis utilized in Eq. \eqref{eq:vecs2}, $|n_H,n_V\rangle$, can be expressed in the single-photon basis $\{|H\rangle,|V\rangle\}$ as
\begin{equation}
    \begin{aligned}
    \label{eq:EqBasis}
	& |n_H,0\rangle=|H\rangle^{\otimes n_H}\\
	& |1,1\rangle =
	\frac{|H,V\rangle+|V,H\rangle}{\sqrt{2}}\\
	&\qquad\vdots\\
        & |n_H,n_V\rangle =	\displaystyle\binom{N}{n_H}^{-1/2} \left(|H\rangle^{\otimes n_H}|V\rangle^{\otimes n_V}+\mathrm{permutations}\right).
    \end{aligned}
\end{equation}
These equivalences allow us to express the symmetric tensor product of qubits, $|q\rangle^{\otimes N}$, in the Fock basis.
This results in a complete correspondence with angular-momentum coherent states, that is
\begin{equation}
    \begin{aligned}
	\label{eq:QNS}
        & |q\rangle^{\otimes N} 
        = \displaystyle\sum_{m=0}^N \binom{N}{m}^{1/2}\alpha_H^{m}\alpha_V^{N-m}|m,N-m\rangle
        =|\vec{s}\rangle,
    \end{aligned}
\end{equation}
where $\alpha_H=\chi$, $m=n_H, \alpha_V=\xi$ and $N-m=n_V$, as one can see by comparison with Eq. (\ref{eq:vecs2}).
Note that Eqs. \eqref{eq:qn} and \eqref{eq:QNS} correspond to the particle and mode description of angular-momentum coherent states, respectively \cite{DJK2015}.

\paragraph{Criteria of nonclassicality.}

To certify nonclassical behavior we consider quasiprobabilities for quantum coherence (QPQC) \cite{SW2018}. 
Such a criterion makes use of the optimal decomposition of the state in terms of elements within a given set of classical states $\{|c\rangle\}$,
\begin{equation}
    \hat{\rho} = \sum_i P_i(c_i)|c_i\rangle\langle c_i|
\end{equation}
Note that a possible additional residual component outside the linear span of classical states does not exist for the notions under study \cite{SW2018, SPDBBS2021}. 
Nonclassicality is then revealed by negativities, $P(c_i)<0$ for some $i$. 
The required stationary states $\{|c_i\rangle\}$, assumed normalized, can be computed by calculating the optimum $\langle c_i|\hat{\rho}|c_i\rangle$. 
For entanglement, where the classical states are the separable ones, this optimization leads to the separability eigenvalue equations \cite{SV2013}
\begin{equation}
    \label{eq:SEE}
    \hat{\rho}_{q^{\otimes(N-1)}} |q_i\rangle=g_i|q_i\rangle,
\end{equation}
where $ \hat{\rho}_{q^{\otimes(N-1)}}= (\langle q|^{\otimes(N-1)} \otimes \hat{1}) \hat{\rho} (|q\rangle^{\otimes(N-1)} \otimes \hat{1})$.
Once Eq. \eqref{eq:SEE} is solved, the quasiprobabilities, $P_i$, can be computed by the linear equation 
\begin{equation}
    \label{eq:PGg}
    \vec{P}=\textbf{G}^{-1}\vec{g},
\end{equation}
where $\vec{g} = [\operatorname{tr} \left(\hat{\rho}\left|c_{i}\right\rangle\left\langle c_{i}\right|\right)]_i$, $\vec P=[P(c_i)]_i$, and $\textbf{G}$ is the Gram-Schmidt matrix, $G_{i,j} = [\operatorname{tr}\left(|c_{i}\rangle\langle c_{i} \mid c_{j}\rangle\langle c_{j}|\right)]_{i,j}$ \cite{BH2012}. 

One strength of this method is that, by substituting the set of classical states, we can study arbitrary notions of quantumness in a similar manner.
Moreover, when two features share the same classical reference, one will always find the same behavior and amount of nonclassicality. 
Because of the one-to-one correspondence of angular-momentum coherent states and product states of identically polarized photons, $|\vec s_i\rangle = |q_i\rangle^{\otimes N}$, nonclassical polarization and entanglement are thus equivalent. 
This further implies that finding $\langle \vec s_i |\hat{\rho}|\vec s_i\rangle \rightarrow \text{``optimal''}$ is equivalent to solving Eq. \eqref{eq:SEE}.

\paragraph{Experiment and state reconstruction.}

In the following, we characterize the nonclassicality of a pair of photons by reconstructing its QPQC from tomographic data. 
The state is produced by type-II spontaneous parametric down-conversion on a waveguide, which generates a two-mode squeezed vacuum state.

Starting with a $774\,\mathrm{nm}$ laser (TiSa), spectrally filtered for a pump width of $1\,\mathrm{ps}$, that pumps the on-chip source, a type-II phase-matched periodically poled potassium titanyl phosphate (ppKTP) waveguide, photon pairs of orthogonal polarizations are generated \cite{ECMS2011,MSPEQDBS2018}.
The large birefringence of the waveguide material results in a temporal delay between signal and idler photons, which is compensated for by placing a \NP{bulk KTP} crystal in the path. 
Next, a polarization tomography of the state is performed \cite{SBKSL2012}.
By introducing a combination of half-wave and quarter-wave plates, it is possible to generate a universal $\mathrm{SU}(2)$ rotation in the polarisation space.
The state is then split by a polarizing beam splitter into two spatial modes. 
Afterward, the detection is performed by a time-bin-multiplexed photon-number-resolved detection system, built from low-loss fibers, 50:50 fiber-splitters, and two high-efficiency (\NP{$>80\%$}) superconducting nanowire single-photon detectors with a resolution up to eight photons \cite{TEBSS2021,PSBS2022}.

\begin{figure}
    \includegraphics[width=0.9\columnwidth]{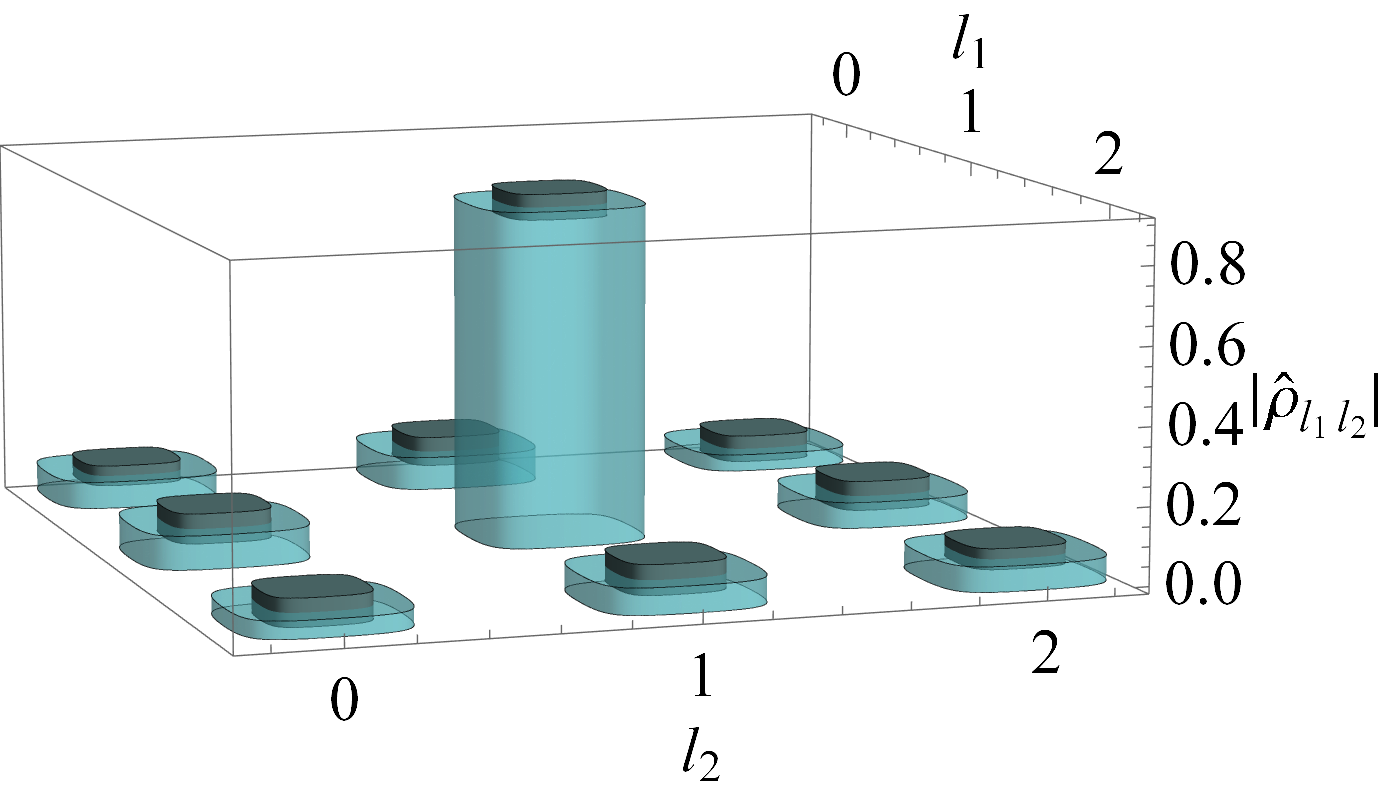}
    \caption{%
        Reconstructed density operator for the two-photon subspace in the two-mode Fock basis.
        The rows and columns are labeled by $l_1$ and $l_2$, respectively.  
        Darker bars show error margins ($\pm 30\sigma$).
    }\label{fig:reconstructeddensity}
\end{figure}

We record in a matrix $C_{n_+,n_-}$ the number of events of photon-number pairs $(n_+,n_-)$. 
For every setting of the half- and quarter-wave plate, $\Omega_j$, one matrix is sampled $C_{n_+,n_-}(\Omega_j)$. 
During the experiment, 156 different configurations were implemented, ensuring a tomographically overcomplete measurement for improving statistical significance.
Importantly, data were not corrected for detector imperfections \cite{SSG2009}.
Thus, all impurities of the system show up as noise in the reconstructed states.
For an $N$-photon subspace, $N=n_+ + n_-$, we determine from the recorded counts the probability of finding $k$ photons in one mode and $N-k$ photons in the other one, resulting in
\begin{equation}
    p_k(\vec \Omega_j)
    =\frac{C_{k,N-k}(\Omega_j)}{\sum\limits_{k=0}^N C_{k,N-k}(\Omega_j)}.
\end{equation}
The density matrix is obtained from these probabilities; details of the underlying inversion can be found in the Supplemental Material \cite{SM}. 
The resulting density matrix for $N=2$, i.e. the two-photon subspace, is shown in Fig. \ref{fig:reconstructeddensity}. 
As expected, the reconstructed state is a good approximation to $|1,1\rangle$.
To estimate the uncertainty, the standard error of the detection probability is propagated via quadratic error techniques.

\paragraph{Analysis of quantumness.}

\begin{figure}
     \centering
     \includegraphics[width=\columnwidth]{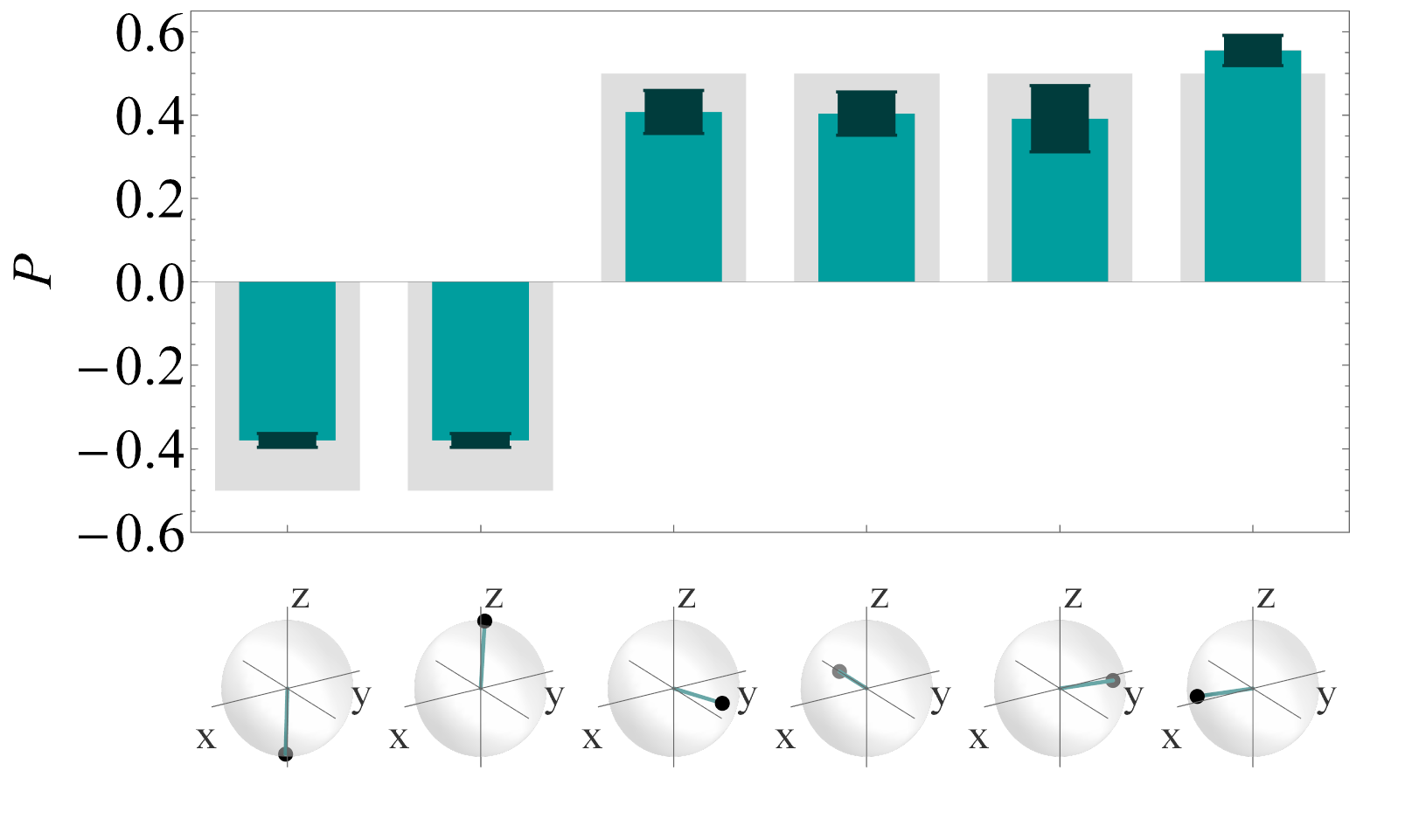}
     \caption{%
        Reconstructed QPQC as a function of the corresponding stationary state. 
        The states required for the decomposition are represented by their position on the Bloch/Poincar\'e sphere. 
        Black bars represent uncertainties ($\pm5\sigma$). 
        In grey, the ideal case $\rho=|1,1\rangle\langle1,1|$ is depicted.
     }\label{fig:QPQC}
\end{figure}

By solving the separability eigenvalue equations, Eq. \eqref{eq:SEE}, for the reconstructed $\hat{\rho}$, the stationary states, $|s_i\rangle = |q_i,q_i\rangle$, and the corresponding quasiprobabilities $P_i$ can be determined.
Two significant negative values of the quasiprobabilities are found, cf. Fig. \ref{fig:QPQC}, which are compared with the ideal case. 
In addition, the QPQC fully describes the state, $\|\hat{\rho} - \sum_{i} P\left(q_i\right) |q_i,q_i\rangle\langle q_i,q_i|\| \approx 6\times 10^{-10}$.
The negativities account for the quantum behavior of the state regarding both polarization nonclassicality and entanglement, implying the state can neither be expressed as a convex combination of separable nor classical polarization states.

\begin{figure}
     \centering
     \includegraphics[width=0.9\columnwidth]{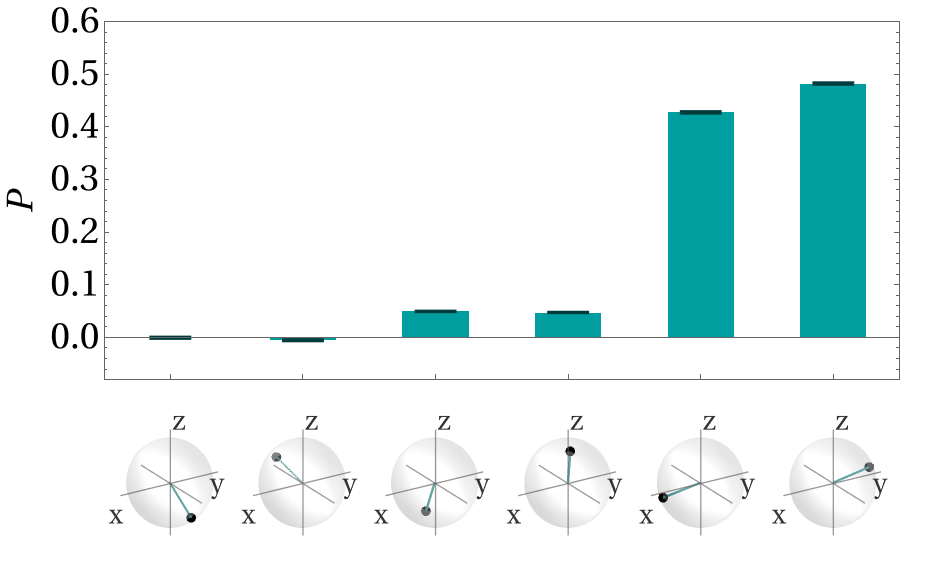}
     \caption{%
        Quasiprobabilities for the reconstructed state of now distinguishable photons, including error bars (black, $\pm5\sigma$).
     }\label{fig:distinguishable}
 \end{figure}

To show the effectiveness of the QPQC negativities as a quantumness signature, we compare the previous result with the reconstruction of a classical state. 
By removing the delay-compensating crystal, distinguishability between photons is introduced, which is tested with polarization-based coincidences \cite{LKNBLGF2018}.
In Fig. \ref{fig:distinguishable} the quasiprobability representation for the scenario under consideration is shown. 
In contrast to the case of highly indistinguishable photons, we find non-significant negativities, effectively being consistent with zeros. 
Thus, the state presents itself as an incoherent mixture of diagonal and anti-diagonal photons, $\hat\rho\approx \frac{1}{2}|D,D\rangle\langle D,D|+ \frac{1}{2}|A,A\rangle\langle A,A|$.
Therefore, our method delivers an appropriate description for distinguishable photons, without falsely indicating polarization nonclassicality and entanglement when they are not present.

A desirable property for the indicator of nonclassicality is the invariance under polarization basis rotations;
since $\mathrm{SU}(2)$ transformations do not alter the polarization structure, they ought not to shift its classical or quantum character \cite{BGB2017}.
From the entanglement perspective, this implies invariance under local transformations, which is also necessary for well-behaved measures of multiparticle entanglement \cite{VPRK1997, SPBS2019}. 
Accordingly, we analyze the dependence of quasiprobabilities on the polarization basis, parametrized by $\theta$.
An $\mathrm{SU}(2)$ operator transforms the reconstructed density matrix as $\hat \rho' = \hat U(\theta)\,\hat \rho\,\hat U(\theta)^{\dag}$, with $\hat U(\theta)^\dag\hat U(\theta) = \hat 1$, mapping stationary states analogously, $|c_i'\rangle=\hat U(\theta)|c_i\rangle$. 
This implies that $\vec{g}$ and $\textbf{G}$ in Eq. \eqref{eq:PGg} remain invariant, $\vec{g'}=\vec{g}$, and $G'_{ij} = G_{ij}$. 
Therefore, the quasiprobabilities in the rotated frame are indeed invariant under polarization basis transformations.

\begin{figure}
     \centering
     \includegraphics[width=0.9\columnwidth]{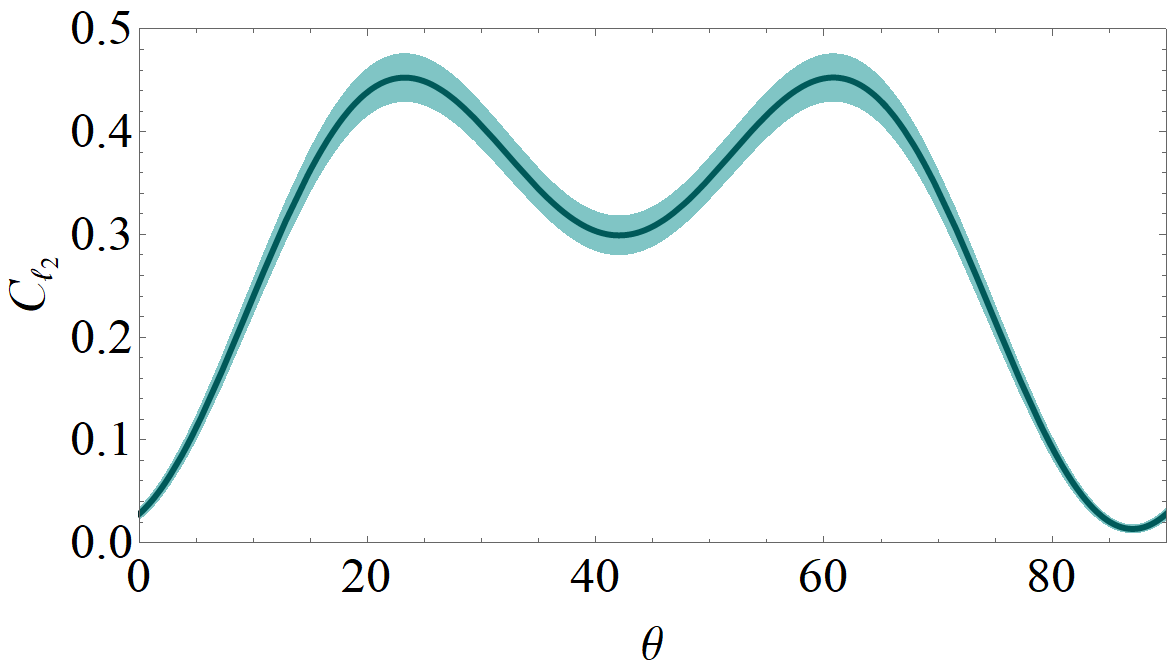}
     \caption{%
        $C_{\ell^2}$ for the reconstructed state (solid line) as a function of the direction of the polarization basis, $\theta$.
        The lighter band represents a $\pm 5\sigma$ uncertainty.
    }\label{fig:Cl2}
\end{figure}

Such invariance can be compared with other measures of coherence, usually being highly base-dependent \cite{SAP2017}. 
As an illustration,  we compute the $\ell^2$-norm of coherence, $C_{\ell^2}=\sum_{i\neq j}|\rho_{i,j}|^2$, of the rotated density matrix.
As shown in Fig. \ref{fig:Cl2}, the state can even become incoherent for certain bases, with $C_{\ell^2}=0.01$ for $\theta= 87^{\circ}$.
Altogether, quasiprobability distributions emerge as a robust and convenient method for signifying nonclassical polarization and entanglement, both theoretically and experimentally.

\paragraph{Beyond two qubits.}

The key relation between separable states of indistinguishable particles and the concept of angular-momentum coherent states can be generalized to $d$-dimensional qudits, $|q\rangle = \sum_{j=1}^{d}\alpha_j|j\rangle$.
Note that, up to this point, we considered $d=2$, as well as $H$ for $j=1$ and $V$ for $j=2$.
The $N$-fold tensor power yields
\begin{equation}
	\label{eq:stateidentity}
	\begin{aligned}
		&|q\rangle^{\otimes N}		
        =\sum_{j_1,\ldots,j_N=1}^{d} \alpha_{j_1} \cdots \alpha_{j_N}|j_1\rangle \otimes \cdots \otimes|j_N\rangle\\
		=&\sum_{\substack{k_0,\ldots,k_d\in\mathbb N:\\k_0+\cdots+k_d=N}}
		\left(\frac{N!}{k_1!\cdots k_d!}\right)^{1/2}
		\alpha_1^{k_1}\cdots\alpha_{d}^{k_d} |k_1,\ldots,k_d\rangle
		=|s\rangle,
	\end{aligned}
\end{equation}
being angular-momentum coherent states for $d>2$.
The Fock state $|k_1,\ldots,k_d\rangle$ in Eq. \eqref{eq:stateidentity} can be expanded in terms of single-particle spaces,
\begin{equation}
    |k_1,\ldots,k_d\rangle 
    = \frac{|1\rangle^{\otimes k_1} \otimes \cdots \otimes |d\rangle^{\otimes k_d}+\text{permutations}}
    {\left(\frac{N!}{k_1!\cdots k_d!}\right)^{1/2}}.
\end{equation}
This abstraction yields generalized $|s\rangle$ for a system of $N$ bosons. 
In the qudit case, $|k_1,\ldots,k_d\rangle$ are generalized Dicke states \cite{H2016}. 
For $d=2$, one recovers Eq. \eqref{eq:EqBasis}.

In general, one can follow an equivalent optimization procedure to find the stationary vectors and their corresponding eigenvalues as described via Eqs. \eqref{eq:SEE} and \eqref{eq:PGg}.

In the case of distinguishable degrees of freedom, it has been shown that the separability eigenvalue equations are invariant under local unitary transformations.
As a corollary when considering high-dimensional, identical qudits and exploiting their resulting symmetry, we recover the invariance under polarization rotations;
the transformed operator in a symmetric scenario, $\hat \rho'=\hat U^{\otimes N}\,\hat \rho\,\hat U^{\dag \otimes N}$, includes accordingly transformed stationary states while leaving quasiprobabilities invariant as previously discussed for $d=2$.

Complementary to the quasiprobability distribution approach, a witnessing strategy can be utilized to certify the entanglement of the state \cite{SV2013}, avoiding a full state reconstruction.
This additional approach is derived in the Supplemental Material \cite{SM} for our unified forms of quantumness.
The latter is based on the expected value of an observable test operator that, for entangled photon-pair states, relates to generalized Dicke states \cite{T07}.
This method proves useful when tackling the separability eigenvalue equation becomes challenging, e.g., in scenarios with a large number of photons.

\paragraph{Conclusion.}

We devise a unified characterization of polarization nonclassicality and entanglement of indistinguishable photons via a single quasiprobability distribution.
By showing that angular-momentum coherent states and symmetric tensor-product states are identical, we derive one quasiprobability for quantum coherence for both quantum features, whose negativities are a necessary and sufficient criterion, and are further supplemented by our witnessing approach.
The theory is directly applied to our experiment, based on a high-performance photon-pair source.
From a tomographic reconstruction, we demonstrate the correlation between the distinguishability and loss of quantumness.
An additional comparison with other notions of quantum coherence shows the polarization-basis independence of our approach.
While many quantum phenomena are studied independently from one another, we believe that finding their common ground---as we did here for two properties---is vital for a fundamental understanding of quantum effects and their exploitation in quantum technologies.

\paragraph{Acknowledgments.}
L.A. and J.S. acknowledge financial support from the Deutsche Forschungsgemeinschaft (DFG, German Research Foundation) through  the Collaborative Research Center TRR 142 (Project No. 231447078, project C10). 
N.P., B.B. and C.S. acknowledge funding from the ERC project QuPoPCoRN (Grant No. 725366).
This work was further funded through the Ministerium f\"ur Kultur und Wissenschaft des Landes Nordrhein-Westfalen through the project PhoQC: Photonisches Quantencomputing.
E.A. acknowledges funding from the Der Wissenschaftsfonds FWF (Fonds zur F\"orderung der wissenschaftlichen Forschung) Elise Richter - HyDRA (V1037).

\bibliography{Main_PRL}

\end{document}


\title{
Supplemental Material
}

\author{Laura Ares}
\affiliation{Theoretical Quantum Science, Institute for Photonic Quantum Systems (PhoQS), Paderborn University, Warburger Stra\ss{}e 100, 33098 Paderborn, Germany}

\author{Nidhin Prasannan}
\affiliation{Integrated Quantum Optics Group, Institute for Photonic Quantum Systems (PhoQS), Paderborn University, Warburger Stra\ss{}e 100, 33098 Paderborn, Germany}

\author{Elizabeth Agudelo}
\affiliation{Atominstitut, Technische Universit\"at Wien, Stadionallee 2, 1020 Vienna, Austria}

\author{Alfredo Luis}
\affiliation{Departamento de  \'Optica, Facultad de Ciencias F\'{\i}sicas, Universidad Complutense, 28040 Madrid, Spain}

\author{Benjamin Brecht}
\affiliation{Integrated Quantum Optics Group, Institute for Photonic Quantum Systems (PhoQS), Paderborn University, Warburger Stra\ss{}e 100, 33098 Paderborn, Germany}

\author{Christine Silberhorn}
\affiliation{Integrated Quantum Optics Group, Institute for Photonic Quantum Systems (PhoQS), Paderborn University, Warburger Stra\ss{}e 100, 33098 Paderborn, Germany}

\author{Jan Sperling}
\affiliation{Theoretical Quantum Science, Institute for Photonic Quantum Systems (PhoQS), Paderborn University, Warburger Stra\ss{}e 100, 33098 Paderborn, Germany}  

\date{\today}

\begin{abstract}
    The nonclassicality that describes a pair of entangled photons is characterized by reconstructing its quasiprobabilities of quantum coherence from tomographic experimental data.
    Supplementary information about the experiment is provided.
    The detailed process of reconstructing multiphoton density operators from the recorded data is described.
    In addition, the methods for the reconstruction of quasiprobabilities and the derivation of entanglement witnesses are outlined.
    Finally, a concise overview of the error propagation is given.
\end{abstract}
\maketitle

\section{Setup and data}
\label{sec:experimentdata}

\begin{center}
\begin{figure}[h]
     \centering
     \includegraphics[width=.98\columnwidth]{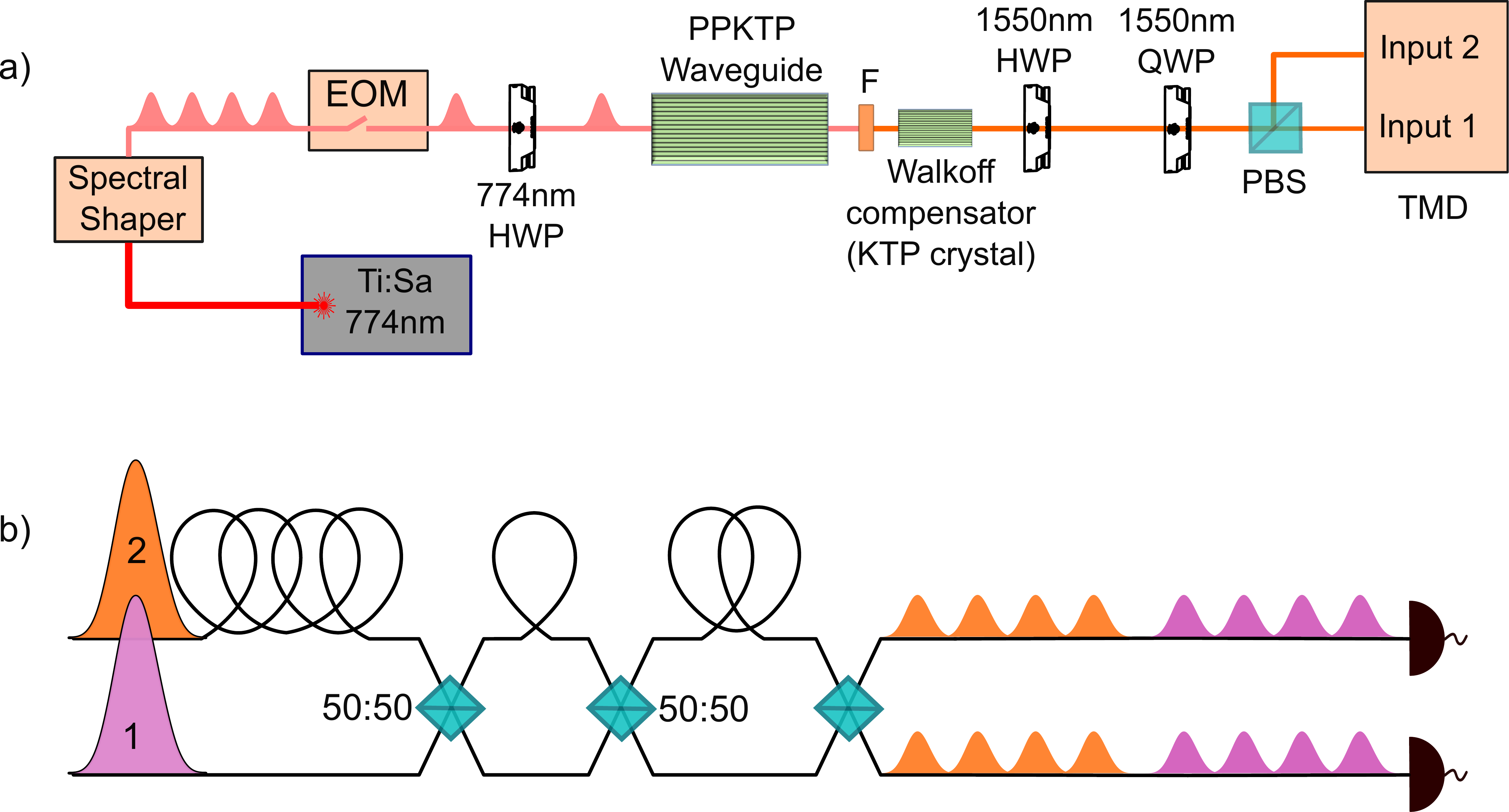}
     \caption{%
     \NP{(a) Sketch of a pulsed PDC source with a two photon state polarization tomographic arrangement (HWP+QWP+PBS). (b) A resource efficient 16-bin time multiplexed detection scheme, designed to resolve eight photons from each tomographic output mode.}}
     \label{fig:setupsketch}
\end{figure}

\end{center}

Here, we describe the experiment \LA{(Fig. \ref{fig:setupsketch})} and the sampled data.
The state under study is produced by type-II spontaneous parametric down-conversion on a waveguide, which generates a two-mode squeezed vacuum state.
Considering the specific conditions of the present experiment, the generated state corresponds effectively to an entangled pair of photons.

\subsection{Additional setup details}

Pump laser pulses of $100\, \mathrm{fs}$ are passing through a folded $4f$ arrangement with a grating and \LA{a} slit.
\NP{\LA{This} provides proper pump bandwidth selection for the phase-matched ppKTP waveguide to do source engineering, resulting a decorrelated PDC state from the source.}
A $78\,\mathrm{MHz}$ electronic signal from the laser is down-scaled to $\approx1\,\mathrm{MHz}$ by a clock synthesizer \NP{\LA{, producing} multiple phase locked clock signals with the down-scaled repetition rate} \LA{that} then drive the whole experiment.
An electro-optic modulator is placed in the laser path to pick pulses with a repetition rate of $\approx1\,\mathrm{MHz}$, based on the reference clock signal.
Such picked pulses produce a sufficiently large time window between the pump pulses\LA{.
This} enables enough \NP{gated time bins in-between every clock-signal (laser pulse) to collect number resolved signal and idler photon counts from our time-bin multiplexing (TMD) detector ouput.
This implies \LA{that the} time-tagger module \LA{and the} pulse\LA{-}picking device are all time synchronized with the clock signal for an efficient data collection.} 
Current TMD designs separate photon bins $100\,\mathrm{ns}$---much higher than detector dead time of $\approx60\,\mathrm{ns}$---apart from each other and provide eight time-bins for each arm (four time-bins for a single input).
We used both available TMD inputs to send signal and idler beams of the parametric down-conversion;
so we get a total of $16$ time bins at the output.
This $16$-bin output is gated by the clock trigger signal, then time-tagged and counted.

Without the TMD network, the source is characterized by selected features.
The setup generates closely degenerated photon pairs with a central wavelength of around $1548\,\mathrm{nm}$ through a spontaneous parametric down-conversion process.
A broad-band filter ($10\,\mathrm{nm}$ full width at half maximum) is used to avoid the side lobes of the phase-matched sine cardinal (sinc) spectrum and provide a decorrelated state with high purity.
Photons in the TMD detection are collected by single-mode fibers and detected by superconducting nanowire single-photon detectors with detection efficiencies around $82\%$.
Overall, the source shows a Klyshko efficiency of $30\%$. 
The indistinguishability of photon pairs is tested with a polarisation-based Hong--Ou--Mandel dip, as described later.
Because of the large birefringence of the waveguide material, orthogonally polarized photons show a temporal delay between them when exiting the chip.
By placing a bulk KTP crystal in the collinear path, the temporal delay is compensated, and a Hong--Ou--Mandel interference visibility of $>91\%$ is observed.
To access many points on the Poincar\'e sphere, quarter- and half-wave plates introduce polarization transformations to move around the Poincar\'e sphere.
For each setting, data are accumulated for $300\,\mathrm{s}$.

\subsection{Data}

\begin{figure}[t]
    \includegraphics[width = .6\columnwidth]{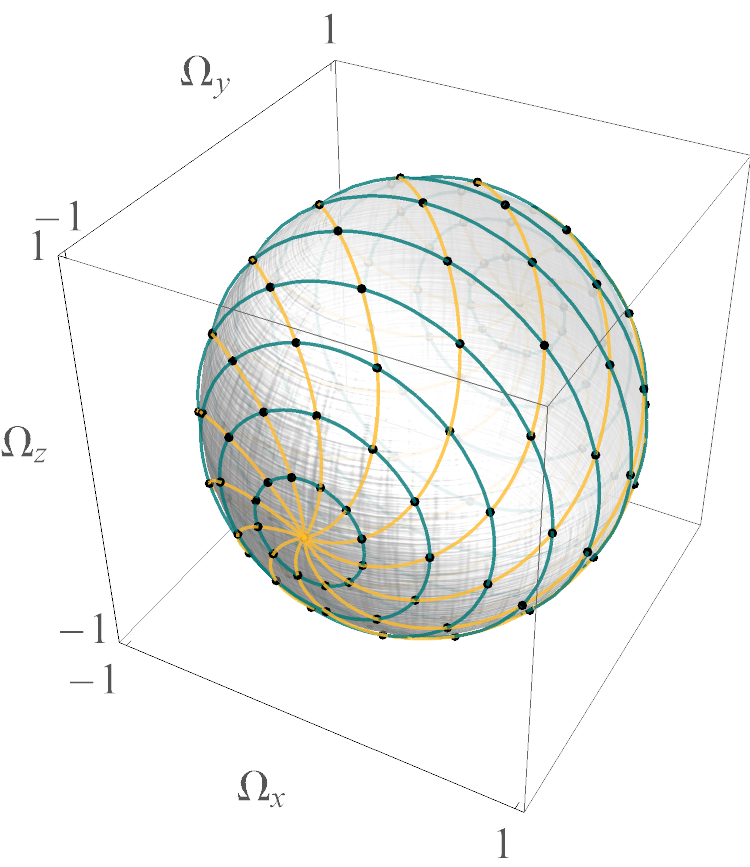}
    \caption{%
        All implemented polarization measurement settings from Eq. \eqref{eq:waveplatesettings} as directions $\vec \Omega$ (black points) on the Poincar\'e sphere.
        Yellow and green curves connect the points with constant half-wave plate angles and quarter-wave plate angles, respectively.
    }\label{fig:polarizationmeasurements}
\end{figure}

The reported results were extracted from two data sets with different pump powers. 
The state with indistinguishable photons was generated with a pump beam of mean photon number $\bar n=0.20314 \left(1\pm5\times10^{-4}\right)$ whereas the state of distinguishable photons was generated with a pump of $\bar n=0.07802\left(1\pm 9\times 10^{-4}\right)$. 
For the subsequent analysis, all registered data were considered;
that is, the values were not corrected for losses \cite{SSG2009}.
The quantum tomography was carried out by measurements with the following waveplate settings:
\begin{equation}
    \label{eq:waveplatesettings}
    \begin{aligned}
        2\theta_\mathrm{HWP}\in&\{0^\circ,15^\circ,\ldots,165^\circ\}\\
        \text{and}\quad
        2\theta_\mathrm{QWP}\in&\{-90^\circ,\ldots,+90^\circ\},
    \end{aligned}
\end{equation}
both linearly spanned with $15^\circ$ steps, yielding $156$ coincidence matrices $C$ per pump power.
This number of settings ensures the relatively dense covering of the entire Poincar\'e sphere, shown in Fig. \ref{fig:polarizationmeasurements},
which corresponds to the directions given by
\begin{equation}
    \vec \Omega=
    \begin{bmatrix}
        \cos(2\theta_\mathrm{QWP})\sin(4\theta_\mathrm{HWP}-2\theta_\mathrm{QWP})\\
        \sin(2\theta_\mathrm{QWP})\\
        \cos(2\theta_\mathrm{QWP})\cos(4\theta_\mathrm{HWP}-2\theta_\mathrm{QWP})
    \end{bmatrix}.
\end{equation}
When restricting to a given total photon number, i.e., $n_++n_-=N$, one obtains the counts $C_{k,N-k}$ for the $N$-photon subspace with $k=n_+$ and $N-k=n_-$.

\begin{figure}[h]
    \includegraphics[width=0.495\columnwidth]{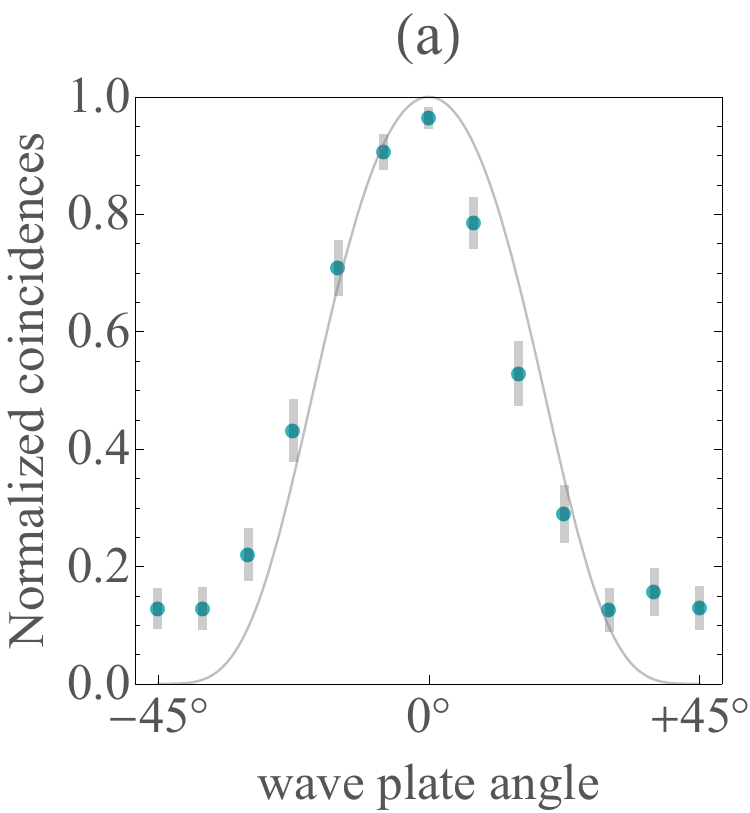}
    \includegraphics[width=0.495\columnwidth]{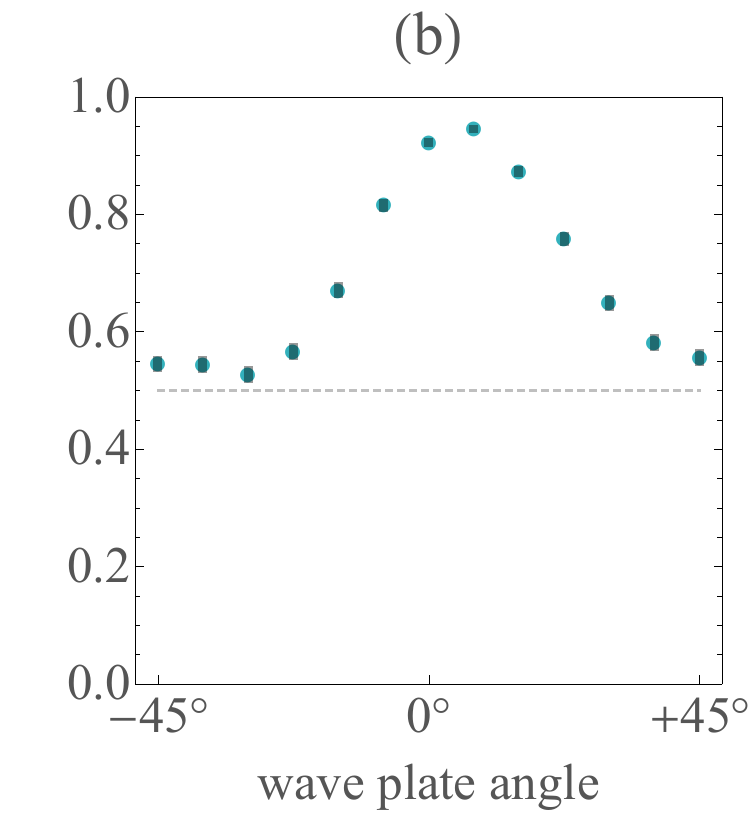}
    \caption{%
        Measured and normalized coincidences as a function of the quarter-wave plate angle $\theta_\mathrm{QWP}$, in the interval $[-45^\circ,+45^\circ]$ with $15^\circ$ steps, and fixed half-wave plate, $\theta_\mathrm{HWP}=0^\circ$.
        Plots (a) and (b) correspond to the indistinguishable and distinguishable cases, respectively.
        Error bars (gray) represent a five-standard deviation ($5\sigma$) error margin.
        The ideal scenario---that is, for perfectly indistinguishable $H$-$V$ photons (solid gray line)---is shown for comparison in (a).
        In (b) the classical threshold of $0.5$ is also indicated (dashed, gray line).
    }\label{fig:HOM}
\end{figure}

To analyze the quality of the generated nonclassically correlated states, we compute the normalized single-photon coincidences in the two-photon subspace, $C_{1,1}/(C_{0,2}+C_{1,1}+C_{2,0})$.
They are displayed in Fig. \ref{fig:HOM} for indistinguishable and distinguishable scenarios in plots (a) and (b), respectively. 
For the ideal generation of horizontally and vertically polarized indistinguishable photons, $(|H\rangle\otimes|V\rangle + |V\rangle\otimes|H\rangle)/\sqrt{2}$, the expected behavior is included as a reference curve.
The coincidence probability can be interpreted in terms of the Hong--Ou--Mandel interference for polarization, with the certainty of coincidences at $0^\circ$ in the ideal case and no coincidences at $\pm 45^\circ$.
Since the ideal state transforms to $(|R\rangle\otimes|R\rangle + |L\rangle\otimes|L\rangle)/\sqrt{2}$ in a right ($R$) and left ($L$), circular polarization basis.
The visibility of the oscillation for the data with the TMD is about $90\%$, certifying the indistinguishability of the photons, necessary for the existence of quantum correlations. 
In addition, in Fig. \ref{fig:HOM} (b), we can see that in contrast to the indistinguishable case, the coincidences for the distinguishable photons never reach a value close to zero.
In fact, they are bounded by the classical threshold of $50\%$ (horizontal line).

\section{Quantum state reconstruction}
\label{sec:statereconstruction}

To estimate the $N$-photon density matrix, one may start calculating the expression for the positive operator-valued measure (POVM) in the Fock representation, considering the probabilities indicated by coincidence counts. 
To derive the values determining the density matrix entries from the latter, we applied a Moore--Penrose inversion procedure to the transformation matrix.
The methods are outlined next.

\subsection{Transformed POVM}

We begin with considering the POVM for no waveplate rotations, i.e., $\vec \Omega = \left[0, 0, 1\right]^T$, that is given by $|n_H,n_V\rangle \langle n_H,n_V|$ for $n_H + n_V = N$ in the Fock representation.
For other orientations, we apply a beam-splitter-type transformation to the polarization modes.
That is
\begin{equation}
    \begin{bmatrix}
        \hat a_+ \\ \hat a_-
    \end{bmatrix}
    \mapsto
    \begin{bmatrix}
        t & r \\ - r^\ast & t^\ast
    \end{bmatrix}
    \begin{bmatrix}
        \hat a_H \\ \hat a_V
    \end{bmatrix},
\end{equation}
with $|t|^2+|r|^2=1$ and annihilation operators, $\hat a$, subscripts $+, -$ representing the outputs of the waveplate transformations, and $H, V$ for the input polarizations.
This rotation relates to the Poincar\'e vector as $\vec \Omega = \left[\sin\theta\cos\phi, \sin\theta\sin\phi, \cos\theta\right]^T$, with $|t|=\cos(\theta/2)$, $|r|=\sin(\theta/2)$, and $\arg r-\arg t=\phi$.
The transformed POVM element $|n_+,n_-\rangle\langle n_+,n_-|$, for each measurement orientation $\vec \Omega$ with $k$ photons in one detector and $N-k$ in the other, corresponds to the POVM element $\hat\Pi_k(\vec \Omega) = |k,N-k\rangle\langle k,N-k|$.
\begin{widetext}
    \noindent
    They are explicitly given by
    \begin{equation}
    \label{eq:transformedfock}
        |n_+,n_-\rangle = |k,N-k\rangle
        = \frac{\left(t\hat a_H+r\hat a_V\right)^{\dag k}}{\sqrt{k!}}
        \frac{\left(-r^\ast\hat a_H+t^\ast\hat a_V\right)^{\dag (N-k)}}{\sqrt{(N-k)!}}
        |\mathrm{vac}\rangle
        = e^{i\varphi}
        \sum_{l=0}^N e^{i\phi l}q_{l,k}\,|n_H=l,n_V=N-l\rangle,
    \end{equation}
    where we have defined the global phase $e^{i\varphi} = \left(\frac{t}{|t|}\right)^{N-k}\left(\frac{r^\ast}{|r|}\right)^{k}$, and
    \begin{equation}
    \label{eq:qparameter}
        q_{l,k}
        = \sum_{u=\max\{0,l-k\}}^{\min\{l,N-k\}} \frac{(-1)^u \sqrt{k!(N-k)!l!(N-l)!}}
        {(l-u)!(k-l+u)!u!(N-k-u)!}
        |t|^{N-k+l-2u}
        |r|^{k-l+2u}.
    \end{equation}
\end{widetext}

\subsection{Reconstructed states}

Let us denote by $\vec \Omega_j$ ($j=1,\ldots, 156$) all measured directions; i.e., the ones depicted on the Poincar\'e sphere in Fig. \ref{fig:polarizationmeasurements}. 
From the recorded counts in the $N$-photon subspace, $C_{k,N-k}$, one can now determine the probabilities of getting $k$ photons in the first detector as
\begin{equation}
    p_k(\vec \Omega_j)=\frac{C_{k,N-k}}{
    \sum\limits_{\substack{ n_+,n_-\in\mathbb N: \\ n_++n_-=N}} C_{n_+,n_-}}.
\end{equation}
In addition, the $N$-photon density operator in the computational basis may be expanded as
\begin{equation}
    \label{eq:densityoperator}
    \hat\rho=\sum_{l_1,l_2=0}^N\rho_{l_1,l_2}|l_1,N-l_1\rangle\langle l_2,N-l_2|.
\end{equation}
Therefore, taking into account the decomposition in Eq. \eqref{eq:transformedfock}, such probabilities can be written in the rotated frame as
\begin{equation}
    \label{eq:measuredprobabilities}
    \begin{aligned}
        p_k(\vec \Omega_j)
        =&\mathrm{tr}\left(
        \hat\rho\hat\Pi_{k}(\vec \Omega_j)
        \right)
        \\
        =&\sum_{l_1,l_2=0}^Ne^{i(l_2-l_1)\phi_j}
        q_{l_1,k}(\theta_j)q_{l_2,k}(\theta_j)\rho_{l_1,l_2}.
    \end{aligned}
\end{equation}

Consequently, we have $\vec p=Q\vec\rho$ in which we have defined the vectors $\vec\rho=[\rho_{l_1,l_2}]_{(l_1,l_2)}$ and $\vec p=[p_k(\vec \Omega_j)]_{(k,j)}$, each element of these vectors is represented in terms of a pair of indices.
In addition, the matrix $Q$ relates the density operator with the measured probabilities as $Q=[e^{i(l_1-l_2)\phi_j}q_{l_1,k}(\theta_j)q_{l_2,k}(\theta_j)]_{(k,j),(l_1,l_2)}$ for all $k$ and $j$ values labelling the elements of $\vec p$.
In order to get the entries of the density matrix, one needs to solve $\vec\rho=Q^{-1}\vec p$.
This was implemented numerically by applying the Moore-Penrose inverse to the rectangular matrix $Q$.
The latter is the most widely recognized generalization of the inverse matrix, and this pseudo-inverse functions like a conventional inverse when $Q$ is invertible \cite{BH2012}. 
The reconstructed density operators represent the basis for the characterization of quantum correlations. 

\section{Reconstructing quasiprobabilities}
\label{sec:quantumcharacterization}

Let us recall that, in terms of the classical basis, quantum states can be represented by quasiprobabilities.
To reconstruct such distributions, one solves the linear equation $\vec g = G\vec p$ \cite{SW2018}. 
To that end, one decomposes the two-photon density operator in terms of Pauli matrices as 
\begin{equation}
    \hat\rho=\sum_{w,w'\in\{0,x,y,z\}}\Gamma_{w,w'}\hat\sigma_w\otimes\hat\sigma_{w'},
\end{equation}
where symmetry implies that $\Gamma_{w,w'}=\Gamma_{w',w}$ holds true.
Furthermore, our qubit state can be decomposed as
\begin{equation}
    \label{eq:qubit}
    |q\rangle\langle q|=\frac{1}{2}\left(\hat\sigma_0+\sum_{w\in\{x,y,z\}}\gamma_w\hat\sigma_w\right),
\end{equation}
with $\vec\gamma=[\gamma]_{w\in\{x,y,z\}}$ being the normalized vector, with $\sum_{w\in\{x,y,z\}}\gamma_w^2=\vec\gamma^\mathrm{T}\vec \gamma=1$, that defines the state on the Bloch sphere.
Then, the reduced density operator $\hat\rho_q$ associated to the separability eigenvalue equation $\hat{\rho}_q|q\rangle=g|q\rangle$ \cite{SV2013} takes the form  
\begin{equation}
    \label{eq:reduceddensity}
    \begin{aligned}
        \hat \rho_q
        &=\left(\sum_{w\in\{0,x,y,z\}}\Gamma_{w,0}\gamma_w\right)\hat\sigma_0\\
        &+\sum_{w'\in\{x,y,z\}}\left(
        \Gamma_{0,w'}+\sum_{w\in\{x,y,z\}}\Gamma_{w,w'}\gamma_w
        \right)\hat\sigma_{w'}.
    \end{aligned}
\end{equation}
Notice that a generic $|q\rangle$ is an eigenvector of a general operator $\hat M$ if and only if the commutator relation $[\hat M,|q\rangle\langle q|]=0$ holds.
For qubits, in particular, where $\hat M=\mu_0\hat\sigma_0+\sum_{w\in\{x,y,z\}}\mu_w\hat\sigma_w$ and $\vec \mu=[\mu_w]_{w\in\{x,y,z\}}$, the vanishing commutator condition is equivalent to $\vec \mu\times\vec \gamma=0$, likewise  $\vec\gamma\parallel\vec\mu$.

In terms of Eqs. \eqref{eq:qubit} and \eqref{eq:reduceddensity}, the condition for parallel vectors reads
\begin{equation}
    \label{eq:parallelconstraint}
    \vec \Gamma+\Gamma\vec\gamma = \lambda\vec \gamma,
\end{equation}
where $\vec \Gamma = [\Gamma_{0,w'}]_{w'\in\{x,y,z\}}$ and $\Gamma =[\Gamma_{w,w'}]_{w,w'\in\{x,y,z\}}$, with a proportionality constant $\lambda$.
Thus, we have
\begin{equation}
    \label{eq:quasiprobvecs}
    \vec\gamma=(\lambda\mathrm{id}_3+\Gamma)^{-1}\vec\Gamma,
\end{equation}
where $\mathrm{id}_3$ denotes the $3\times3$ identity matrix.
The required normalization, i.e., $\vec\gamma^\mathrm{T}\vec\gamma=1$, can be recast into the form
\begin{equation}
    \label{eq:sixthorderpolinomial}
    0=\left[\vec \Gamma^\mathrm{T}(\lambda\mathrm{id}_3+\Gamma)^{-2}\vec\Gamma-1\right]\det(\lambda\mathrm{id}_3+\Gamma)^2.
\end{equation}
Herein, the determinant factor is used to obtain a sixth-order polynomial in $\lambda$, which only depends on the coefficients $\Gamma_{w,w'}$ of the density matrix (recall that $\det(\hat M)\hat M^{-1}$ is the adjugate of $\hat M$ which is a function of minors of $\hat M$, thus polynomial).
The solutions $\{\lambda_i\}_{i=1,\ldots,6}$ of Eq. \eqref{eq:sixthorderpolinomial} can be put into Eq. \eqref{eq:quasiprobvecs}, from which one obtain{s $\{|q_i\rangle \langle q_i|\}_{i=1,\ldots,6}$ (likewise, normalized $|s_i\rangle=|q_i\rangle \otimes |q_i\rangle$) and
\begin{equation}
    g_i=\langle s_i|\hat \rho|s_i\rangle.
\end{equation}
Finally, this allows us to compute the quasiprobabilities via $\vec{P}=\textbf{G}^{-1}\vec{g}$.
These steps of constructing the solutions are applied to the reconstructed density operator $\hat\rho$ to determine quasiprobabilities from experimental data.

\subsection{Witnessing approach}
\label{subsec:witnessing}

Complementary to quasiprobabilities is witnessing, which can also verify quantum properties.
This method proves powerful when tackling the separability eigenvalue equation becomes challenging, e.g. in scenarios with a large number of photons.
Exceeding the $N=2$ quasiprobability approach from the letter, we here also derive the methodology required to jointly witness polarization nonclassicality and entanglement and apply it to examples.

The present approach is based on the observation that, for any observable $\hat L$, the expectation value for a classical state, $\hat\rho=\sum_i p_i|s_i\rangle\langle s_i|$ with $p_i\geq0$ for all $i$, is bounded as
\begin{equation}
    \label{eq:classicalconstraint}
    \langle\hat L\rangle=\mathrm{tr}(\hat\rho\hat L)\leq g_{\max},
\end{equation}
where $g_{\max}$ denotes the maximal expectation value for classical states.
Because of convexity, the maximal expectation value for classical states is obtained by pure ones.
Thus, in order to find $g_{\max}$, one solves $\langle s|\hat L|s\rangle\to\mathrm{max.}$, which is the kind of optimization problem discussed in the main part.
Here, only the maximal solution is required to formulate the witnessing criteria in Eq. \eqref{eq:classicalconstraint}.
Given the above considerations, it can be convenient to define a proper witness operator as
\begin{equation}
    \hat W=g_{\max}\hat{\mathbbm 1}^{\otimes N}-\hat L,
\end{equation}
which must have a nonnegative expectation value for classical states, $\langle\hat W\rangle\geq0$, while a negative expectation value certifies the state's nonclassicality.

Note that in the case of entangled photon pairs, one may use Dicke states to witness entanglement, considering $\hat L=|D^N_k\rangle\langle D^N_k|$, where
\begin{equation}
    \begin{aligned}
         &\left|D_{0}^{N}\right\rangle  =|V\rangle^{\otimes N} \\ 
         &\left|D_{1}^{N}\right\rangle  =\frac{|H\rangle \otimes|V\rangle^{\otimes N-1}+\cdots+|V\rangle^{\otimes N-1} \otimes|H\rangle}{\sqrt{N}} \\ & \quad\vdots \\ 
         &\left|D_{N-1}^{N}\right\rangle  =\frac{|V\rangle \otimes|H\rangle^{\otimes N-1}+\cdots+|H\rangle^{\otimes N-1} \otimes|V\rangle}{\sqrt{N}} \\ 
         &\left|D_{N}^{N}\right\rangle  =|H\rangle^{\otimes N}.
     \end{aligned}
     \label{eq:Dickes}
\end{equation}	
Next one proceeds to solving the separability eigenvalue equation $\hat L_{q^{\otimes(N-1)}} |q\rangle = g |q\rangle$, for the $|H\rangle$ and $|V\rangle$ component, i.e., $\langle H|\hat L_{q^{\otimes N-1}}|q\rangle=g\langle H|q\rangle$ and $\langle V|\hat L_{q^{\otimes N-1}}|q\rangle=g\langle V|q\rangle$.
Then, we obtain the following equations:
\begin{equation}
    \label{eq:witnessexample}
    \begin{aligned}
        \frac{k}{N}\binom{N}{k}(|\alpha|^2)^{k-1}(|\beta|^2)^{N-k}\alpha=&g\alpha,
        \\
        \frac{N-k}{N}\binom{N}{k}(|\alpha|^2)^{k}(|\beta|^2)^{N-k-1}\beta=&g\beta,
    \end{aligned}
\end{equation}
with $|\alpha|^2+|\beta|^2=1$.
Observe that this can be readily seen by considering the Fock basis formulation of the optimization,
\begin{equation}
    g=\frac{\langle s|k,N-k\rangle\langle k,N-k|s\rangle}{\langle s|s\rangle}=\binom{N}{k}\frac{|\alpha|^{2k}|\beta|^{2(N-k)}}{(|\alpha|^2+|\beta|^2)^N},
\end{equation}
whose gradient (when set to zero) with respect to $\alpha^\ast$ and $\beta^\ast$ yield the above equations for both components.
The nontrivial solutions ($g\neq0$) of Eq. \eqref{eq:witnessexample} can be straightforwardly obtained via the quotient of both relations, which grants $|\alpha|^2 = \tfrac{k}{N}$ and $|\beta|^2 = \tfrac{N-k}{N}$.
This results in the maximal expectation value for classical states given by
\begin{equation}
    g_{\max}=\binom{N}{k}\frac{k^k(N-k)^{N-k}}{N^N},
\end{equation}
for $k\in\{0,N\}$ and where one sets $0^0=1$.
In Fig. \ref{fig:witness}, the values of $g_\text{max}$ for different $N$ and $k$ are provided.
Therefore, once the test operator $\hat{L}$ is established, one can compare the corresponding maximum value allowed for classical states to the experimentally determined $\langle \hat{L}\rangle$ and conclude whether the state under study is entangled or not.

\begin{figure}[h]
    \begin{tabular}{p{0.8\columnwidth} p{0.2\columnwidth}}
        \vspace{0pt} \includegraphics[width=.7\columnwidth]{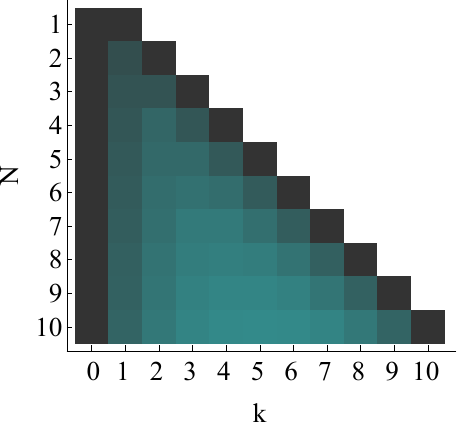}
        & \vspace{-3pt} \includegraphics[width=.12\columnwidth]{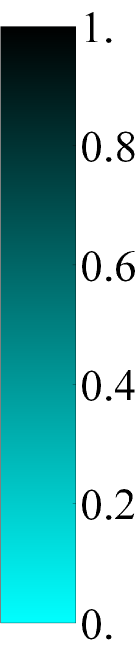}
    \end{tabular}
    \caption{%
        Separable bound $g_{\max}$ for the expectation values of the operators $\hat L = |D^N_k\rangle\langle D^N_k|$, for $0\leq k\leq N$.
        For $k=0$ and $k=N$, we have $g_{\max}=1$ as expected since the corresponding Dicke states are factorizable.
    }\label{fig:witness}
\end{figure}

\section{Remarks on error estimation}
\label{sec:additionalremarks}

We utilize standard error estimations for photo-counting data and quadratic ($p=2$) error propagation methods for the error associated with the density matrix elements,
\begin{align}
    &\sigma(c_k)=\sqrt{\frac{c_k(1-c_k)}{\sum_{k=0}^N c_k}},\quad\text{and}
    \\ \sigma(&F)=\sqrt[p]{\sum_k \left|\frac{\partial F}{\partial c_k}\right|^p\sigma(c_k)^p},
\end{align}
respectively.

In contrast, as the computation of the quasiprobabilities involves solving nonlinear equations, a Monte Carlo error estimation approach is applied.
Let us recall that $\hat\rho$ is the reconstructed density operator and $\sigma(\hat\rho)$ contains the obtained uncertainties of this reconstruction.
Then, via the Monte Carlo simulation, a sufficiently large sample $\{\hat\rho_z\}_{z=1}^Z$ of $Z=30\,000$ density operators is generated.
The distribution of the sample is a multivariate Gaussian such that the sample mean and standard deviation correspond to $\hat\rho$ and $\sigma(\hat\rho)$, respectively.
The quasiprobability representation is computed for each $\hat\rho_z$ separately, e.g., resulting in a quasiprobability $p_{i,z}$.
Then, the mean and standard deviation of all sample elements $\{p_{i,z}\}_{z=1}^Z$ determines the value and uncertainty of the reconstructed $p_i$.

\bibliography{supplement}